\magnification = \magstep1
\pageno=0
%\nopagenumbers
\hsize=15.0truecm
\hoffset=3.0truecm
\hoffset=0.5truecm
\vsize=22.5truecm
%\voffset=2.5truecm
%\voffset=0.0truecm
%\voffset=1.5truecm

\output={\plainoutput}
\pretolerance=3000
\tolerance=5000
\hyphenpenalty=10000    %so as not to break words
\newdimen\digitwidth
\setbox0=\hbox{\rm0}
\digitwidth=\wd0

\def\footnoterule{\kern-3pt \hrule width \hsize \kern 2.6pt
\vskip 3pt}
\def\cl{\centerline}
\def\ni{\noindent}

\def\vs{\vskip 11pt}

\def\solar{\ifmmode _{\mathord\odot}\else $_{\mathord\odot}$\fi}

\font\ksub=cmsy7
\def\teff{T$_{\kern-0.8pt{\ksub e\kern-1.5pt f\kern-2.8pt f}}$}
\font\small=cmr7
%\font\ktitle=cmsy12
%
\vs\vs\vs
\vs\vs\vs
\vs
\vs
\vs
\vs
\vs
\cl{\bf ELEMENTARY PHYSICS}
\vs
\cl{\bf IN THE CELLULAR AUTOMATON UNIVERSE}
\vs\vs
\cl{Robert L. Kurucz}
\cl{Harvard-Smithsonian Center for Astrophysics}
\vs
\vs
\cl{September 7, 2004}
\vs
\cl{Festschrift for my Sixtieth Birthday}
\vs
\cl{Revised, May 18, 2006}
\vs
\vs\vs\vs 
%\cl{DRAFT}
\eject
\vs\vs
\cl{\bf ELEMENTARY PHYSICS}
\vs
\cl{\bf IN THE CELLULAR AUTOMATON UNIVERSE}
\vs\vs
\cl{Robert L. Kurucz}
\vs
\cl{Harvard-Smithsonian Center for Astrophysics}
\cl{60 Garden St, Cambridge, MA 02138}
\vs
\centerline {Abstract}

General relativity is a mathematical model that uses sophisticated geometry 
to describe simple physics.  It agrees with experiment in the few tests
that can be made, but the whole edifice is not physics.  Instead of using 
observations to test that model, I derive a simple empirical model of 
elementary physics and cosmology from the observations.
The observations imply that the universe is a finite cellular automaton; 
that there is no curved space; that fundamental particles are massless; 
that ``massy" particles, including electrons, are composed of fundamental 
particles; that gravitational mass is inertial mass; that black holes are
made from neutrons compressed into bosons; that the universe was produced 
from cold compressed particles, not radiation; and that the universe is not 
expanding.
\vs
\vs
\centerline {\bf GRAVITY}

A physical ``law" on which people seem to agree is that the speed of light 
is constant to all observers, that massless particles move at the speed of 
light in vacuum, and that massy particles have the speed of light as an
unreachable upper limit to their velocities.  This should be the starting point 
of our physics instead of the customary historical development.  The law implies 
that physics is ``digital" instead of ``analog" and that the speed of light is 
some kind of unit.  The easiest explanation is that the universe is a cellular 
automaton$^{\dagger}$ and that the speed of light is the ratio of the space 
step to the time step.  At every clock tick everything moves or shifts one step.  
There are no such things as continuity, or points, or singularities, or infinity 
in physics.  These are mathematical concepts that are used in mathematical 
modelling of physics, but they are not physics.  
 
     Any fundamental particle must move at the speed of light because that is 
the only speed possible in the cellular automaton, one space step per tick, and 
therefore a fundamental particle must be massless.

\ni-----------

\ni $^{\dagger}$A cellular automaton is a lattice of discrete identical sites,
each taking on a finite set of values.  The values evolve in discrete time
steps according to rules that specify the value of each site at the next time 
step in terms of the values of neighboring sites now (cf. Wolfram 1994).  This
paper is not based on the work of Wolfram or of anyone else.  The first draft
was written in January 1975.
\vfill
\eject

     Composite particles are combinations of fundamental particles.  Since
massy particles move at less than the speed of light, massy particles must 
have an internal structure of massless particles that move at the speed of 
light but that execute internal motions in the massy particles so that the 
net velocity is less than the speed of light.  The need to perform internal 
motions prevents a massy particle from ever moving as fast as the speed of 
light no matter how much it is accelerated.

     If massy particles are made from massless particles, then mass does not
exist.  What we perceive as mass is just inertial mass equal to the momentum
divided by the velocity.  All particles have inertial mass.

     If massy particles are made from massless particles, then gravity cannot
depend on mass, but instead on inertial mass, which is identical to 
gravitational mass, and is equal to the momentum divided by the velocity.
All particles have gravitational mass.  All particles are affected by gravity
and attract each other.  

     If a plausible minimum time in particle physics is $\sim$10$^{-28}$s,  
the cell size in the cellular automaton would be $\sim$3$\times$10$^{-18}$ cm, 
about 10000 cells in a classical electron radius.  Discussion of a cellular 
automaton model of the universe is given at the end of this paper.  At low 
resolution, down to nuclear dimensions, the special relativity model of physics 
works well with continuous space-time and continuous Lorentz transformations.
However as v/c approaches unity at very high energies the granularity must 
become apparent in all relativistic effects.

     Since the universe is filled with particles and massive bodies, no particle
or body can travel in a straight line for any significant length of time.
Particle paths are always longer than they seem naively.  Massless particles 
may seem to be slowed by gravity in traveling between two points but the massless 
particles always travel at the speed of light over a longer path.  If a
massless particle is passing through plasma, or gas, or liquid, or a solid,
its velocity does not change; its path wobbles on a microscopic scale so that
the distance traveled is greater.

General relativity is a mathematical model that uses sophisticated geometry to 
describe simple physics.  It is not physics.  There is no such thing as curved 
space.  Space is not a physical concept since the cellular automaton is complete 
in itself.  Many real results of general relativity can be obtained using simple 
empirical arguments.  Below are the standard observations 
from which the properties of gravity can be derived.  Calculations can be made 
with a continuous mathematical models because the scales are so large.  After
the discussion of gravitation, additional sections on particle physics and 
cosmology are meant to suggest the new physics required.
As I am not capable of doing it, I leave it to the readers 
to work out the cellular rules for their favorite particles and to correct 
my guesses about the cellular automaton structure.

\vs\vs
\centerline {The Precession of the Perihelion, or The Deflection of a ``Massy" Particle}
\vs
 Mercury orbiting the Sun is the standard example where the observed 
precession is 13.489"/orbit mostly from precession of the observer's 
coordinate system and from interactions with other bodies in the
solar system (cf. Misner, Thorne, and Wheeler 1973, p.1113).  Only 
0.104"$\pm$0.002"/orbit of the precession is produced by integration over the volumes 
of Mercury and the Sun and non-radial relativistic effects (Shapiro et al. 1972).  
But since we do not yet know the internal structure of Mercury (or the Sun), the 
precession of Mercury is not a test of relativity.  In fact the density 
distribution in Mercury can be estimated by requiring that it produce most
of the observed precession.  In distance, the precession of the perihelion 
is about 29 km/orbit.

     The gravitational field produced by a composite particle or larger body 
varies with velocity and has the same properties as the electric field produced 
by a moving electric point charge that is described in electricity and magnetism 
textbooks.  From Resnick (1968) substituting GM for q/4$\pi \epsilon_{0}$,
\vskip 5pt
$ \vec{f} =  (1-\beta^{2})/[1-\beta^{2} \sin^{2} \alpha]^{3/2} \ GM\hat{r}/r^{2}$,\hskip 150pt [1]
\vskip 5pt
\ni where $\beta = v/c, v$ is the particle velocity, c is the 
speed of light, $\alpha$ is the angle between $\hat{v}$ and
$\hat{r}$, G is the gravitational constant, and 
M = M$_{0}$/(1-$\beta^{2})^{1/2}$ is the gravitational mass.  
The field moves away from the poles, the direction of motion, toward the
equator as the velocity increases.  The $\alpha\beta$ factor is $(1-\beta^2)$
at the poles which goes to 0 as $v$ approaches c.  The $\alpha\beta$ factor 
is $1/(1-\beta^2)^{1/2}$ at the equator which becomes large as $v$ approaches c.
Substituting 
sin$^{2}$ = 1--cos$^{2}$ = 1--$(\hat{v}\cdot\hat{r})^{2}$,
\vskip 5pt
$ \vec{f} =  (1-\beta^{2})/[1-\beta^{2}(1-(\hat{v}\cdot\hat{r})^{2}]^{3/2} \ GM\hat{r}/r^{2}$. \hskip 139pt [2]
\vskip 5pt
The integral of the force over the volumes of Mercury and the Sun does not 
degenerate into a two-body problem but it can be treated as a two-body problem in 
heliocentric coordinates with perturbations.  The subscripts {\small S} and {\small M}
refer to the Sun and Mercury.  
\vskip 5pt
$ \vec{F} =  - G M_M (M_{\odot}+M_M)(1-v^2/c^2)/[1-v^2/c^2(1-(\hat{v}\cdot\hat{r})^{2}]^{3/2} \ \hat{r}/r^{2}$, \ \ \ \ \ \ \ [3]
\vskip 5pt
\ni where $\hat{r}$ and $\hat{v}$ are the position and velocity of the center of mass 
of Mercury relative to the position and velocity of the center of mass 
of the Sun, which are defined to be 0.  The two-body force is purely radial.  The 
acceleration toward the Sun is $\vec{a} = \vec{F}/M_M$.  Typical values in this 
problem are:
period  88 days, r = 0.387 AU or 58 million km; $v$ = 48 km/s; $\beta$ = 0.00016;
$1-\beta^2$ = 0.999999974; M$_{0M}$ = 3.3$\times$10$^{26}$ g; M$_{0M}$/M$_{\odot}$ = 0.000000166;
M$_M$ = 1.0000000128$\times$3.3$\times$10$^{26}$ g; $\hat{v}\cdot\hat{r}$ = -0.20 to +0.20. 

The perturbative forces are defined at the same time and positions as the 
two-body force, at the center of mass of the Sun and the center of mass of Mercury.
Define $\vec{r}_S$ and $\vec{v}_S$ as the position and velocity 
vectors of a mass element in the sun relative to the center of the Sun 
and $\vec{r}_M$ and  $\vec{v}_M$ are the position and velocity 
vectors of a mass element in Mercury relative to the center of mass of Mercury.
Let $\vec{d} = \vec{r} + \vec{r}_S + \vec{r}_M$ and
$\vec{w} = \vec{v} + \vec{v}_S + \vec{v}_M$.
Mercury and the Sun are far apart so they present small solid 
angles to each other.  Assume that the Sun is spherical with radially varying 
density ranging from 0 to 148 g/cm$^3$.  Assume that the Sun rotates 
with a surface equitorial velocity of about 2 km/s and that internal motions 
are smaller than 2 km/s and symmetric about the equator.  In the solar part of 
the integrand of the force, angular effects and retardation effects are small
(which I have tested by numerical integration) so that  $\vec{r}_S$ and 
$\vec{v}_S$ can be ignored and the solar part of the integrand can be factored 
out.  Assume that Mercury is spherical with radially varying density.  The 
rotation of Mercury is small and internal motions are small or 
non-existent so $\vec{v}_M$ is small and  $\vec{w} = \vec{v}$.  The radius 
is 2440 km and the density varies from about 3 to more than 9 g/cm$^3$.  
The force reduces to
\vskip 5pt
\ni $\vec{F}=-GM_M(M_{\odot}+M_M)\int_M (1-v^2/c^2)/[1-(v^2/c^2)(1-(\hat{v}\cdot\hat{d})^{2}]^{3/2} \rho_M \ \hat{d}/d^2 dV_M/M_M$

\hskip 325pt [4]

\ni where $\vec{d} = \vec{r} + \vec{r}_M$ and where 
all the mass points are retarded to the center of mass of Mercury.  
The retardation in time is $dt = -(d-r)/c$, and in position is $\vec{d}\ ' = \vec{d}-\vec{v} (d-r)/c$.
This force has non-radial components.
Since $\beta$ is small the denominator can be expanded to yield
\vskip 5pt
$ \vec{F} = -GM_M(M_{\odot}+M_M)\int_M [\small 1+${\small 1/2}$\beta^{2}-${\small 3/2}$
\beta^{2}(\hat{v}\cdot\hat{d})^{2}]\ \rho_M \ \hat{d}/d^{2} \ dV_M/M_M.$ \ \ \ \ [5]
\vskip 5pt
I approximately computed the precession as follows:  First the orbit was computed 
as a two-body problem with [3] using Butcher's 5th order Runge-Kutta method following 
Boulet (1991) for 50000 steps/day.  The perihelion was determined by 6-point
Lagrangian differentiation.  The two-body orbit repeated the 
perihelion to 100 $\mu$arcsec.  A vectorial correction factor to the two-body 
force for the volume of Mercury was tabulated five times per day for that orbit by 
integrating [5] over 500 density shells, 500 latitudes, and 1000 longitudes 
for a range of Mercury models.   The precession for a uniform density of 
5.426 g/cm${^3}$ was 0.117"/orbit.  An iron-core-with-frosting model that I made 
up inspired by a figure in Schubert et al. (1988) yielded a precession of 
0.107"/orbit.  Thus, using only special relativity, a simple model reproduces 
the precession to within 3 percent. I tried ad hoc fixes to the model until a 
precession of 0.105"/orbit was achieved in agreement with observation.  Remember
that other small volume effects in Mercury and the Sun were ignored in this
calculation; neither is actually spherical.  Seismic 
measurements on Mercury itself will eventually allow a real model to 
be determined that will test the relativistic calculation.
\vs 
\vs
\centerline {The Deflection of Starlight, or The Deflection of a ``Massless" Particle}
\vs
The gravitational field expands from a particle at the speed of light.  
But a ``massless" particle moves at the speed of light.  In the direction
of motion there can be no gravitational field.  In analogy with the relativistic
contraction of the field of a massy particle described above, the whole 
gravitational field of a massless particle is in the plane perpendicular 
to the direction of motion.  Instead of filling a solid angle 4$\pi$, the
field contracts to the ``width" of the equator $d\phi$ with solid angle 
2$\pi d\phi$ and ``strength" $S$.  Then $S2\pi d\phi$ = 4$\pi$ and $Sd\phi$ = 2$\delta_{\perp}$
where $\delta_{\perp}$ is a $\delta$-function in the equatorial plane. The gravitational 
mass of a photon, or other massless particle is 2$\delta$$_{\perp}$E/c$^{2}$.
The gravitational mass averaged over all directions is $E/c^{2}$ and the momentum
of a massless particle is $(E/c)\hat{c}$.
\vs

A massless particle gains energy, blueshifts, as it approaches a massive
body B and loses energy, redshifts, as it departs.  The gain in energy is also 
a gain in mass.  The gravitational mass $M_{E}$ is $E/c^{2} = (1+GM_B/r/c^2)E_{r=\infty}$.

The two-body force on a massless particle passing a massive body B in B-centric 
coordinates is 
\vskip 5pt
$ \vec{F} = -GM_BM_E \ 2 \sin \alpha \ \hat{p}/r^2$ \hskip 200pt [6]
\vskip 5pt
$ \vec{F} = -GM_BM_E \  2 \sqrt{1-(\hat{c}\cdot\hat{r})^2} \ \hat{p}/r^2$
\vskip 5pt

\ni where $\hat{r}$ is the vector from the center of mass of B to the massless 
particle, $\alpha$ is the angle between $\hat{r}$ and $\hat{c}$, and 
$\hat{p}$ is perpendicular to $\vec{c}$, $\hat{c}\cdot\hat{p} = 0$, and lies 
in the plane defined by $\vec{c}$ and $\vec{r}$.  The acceleration is $\vec{a} = \vec{F}/M_E$.
The general case for an extended body B with internal motion is 
\vskip 5pt
$ \vec{F} =  -GM_E \ 2\int_B \rho \sqrt{1-(\hat{c}\cdot\hat{d})^2} \ \hat{p}/d^2 \ dV$. \hskip 140 pt [7]
\vskip 5pt
\ni where $\vec{d} = \vec{r} + \vec{r}_B$ and the mass points are retarded 
to the center of mass of B, $\hat{r}_B$  and $\vec{v}_B$ 
are the position and velocity vectors of a mass element in B relative to the 
position and velocity of the center of mass of B which are defined to be 0.
Here $\hat{p}$ is perpendicular to $\vec{c}$, $\hat{c}\cdot\hat{p} = 0$, 
and lies in the plane defined by $\vec{c}$ and $\vec{d}$.  The retardation in 
time is $dt = -(d-r)/c$, and in position is $\vec{d}\ ' = \vec{d}-\vec{v}_B (d-r)/c$.

     There are no observations of deflection by point masses.  The observations 
are of lensing by ill-defined bodies far away and lensing by the 
sun.  The solar observations are that photons from a distant star or planet are 
deflected by 1.751$\pm$0.002 $R_{\odot}/R_{min}$ arcsec (Robertson, Carter, and 
Dillinger, 1991) as they pass the Sun and are observed at the earth, where $R_{min}$ 
is the  closest approach of the photon to the center of the Sun.  There are also
observations that show that the pathlength for radar echos from the planets to 
the Earth increases when the path passes near the Sun relative to when it is
further away from the Sun.  Shapiro et al. (1971) measured a maximum effect of 
about 180 $\mu$s or 60 km for Earth-Venus-Earth radar echos.

     I approximately computed the deflection as follows:  The 
orbit was computed
as a two-body problem with [7] using Butcher's 5th order Runge-Kutta method 
following Boulet (1991) with the constraint that $v = c$.  The time step was 
0.005 s near the sun and increased outward, and the path ran from -42 million 
km at Venus to +150 million km at the Earth passing the limb of the Sun 
at 700000 km.  The Sun was modelled as a sphere with no internal motions, so no 
retardation.  The radial density distribution was interpolated from Lebretton 
and Dappen (1988), normalized to the solar mass, and the force was integrated 
over 500 density shells, 250 latitudes (in a hemisphere), and 1000 longitudes.   
I integrated the force over the volume at each step and corrected the two-body 
force.  The deflection was found to be 1.751 $R_{\odot}/R_{min}$ arcseconds 
in agreement with observation.  Actually the same deflection is found by 
treating the Sun as a point mass, but the force is slightly weaker near the 
Sun and slightly stronger far away.  Thus the deflection of photons is an 
effect of special relativity.

     The time delay of radar echos can be explained without general 
relativity.  The path for Earth-Venus-Earth radar is rather complicated.  
A beam of small solid angle leaves the radar on the surface of the Earth.  
The beam travels about 640 light seconds, 192 million km, and is deflected 
1.751 $R_{\odot}/R_{min}$ arcseconds by the Sun along the way.  At Venus the 
deflection is 1629 $R_{\odot}/R_{min}$ km compared to a radius of 6070 km. 
The beam hits Venus and is backscattered by roughness on the surface integrated 
over the area of the beam as a function of time.   A small amount of 
backscatter travels back to the radar on Earth and its path is deflected by 
1.751 $R_{\odot}/R_{min}$ arcseconds in passage.  Meanwhile the radar 
receiver on Earth has orbited and rotated for 1280 s, about 40000 km, 
which is about 40 arcseconds seen from Venus.  This round trip yields a 
measurement of the elapsed time to the detection of the beginning of the 
return pulse out of the noise.  Consider a
hypothetical measurement where there is no deflection, and where the beam 
hits the exact center of the disk of Venus and is circularly symmetric.  The
backscatter to the radar on Earth will measure a time and distance that are 
a bit longer than the distance to the centerpoint.  These hypothetical 
times are less than the actual observed times.  The conclusion is that the real
measurement is an integration over area that delays the detectable onset of the 
return pulse by up to 180 $\mu$s or 60 km round trip.  The effective surface 
includes enough of the hemisphere to be up to 30 km beyond the centerpoint.

\vs
\vs
\centerline {Orbits of Binary Stars}

     Binary pulsars are assumed to lose energy through gravitational radiation
(Weisberg and Taylor 1984) and to be a test of general relativity.  However, there are  
energy loss mechanisms that can be treated only empirically or semiempirically 
at the present time that would affect the period.  Because of the additional
effects, observations that match theory perfectly make me very uneasy.

     A neutron star can be crystallized or partially crystallized so that
the neutron spins are aligned and the star becomes piezomagnetic.  In a binary
there is a piezomagnetic tide that varies the magnetic field and generates
low frequency electromagnetic radiation.  There is also a tide in the plasma
around the stars that generates radiation.  In fact the universe should be 
flooded with low frequency electromagnetic radiation from neutron stars
in binaries.             

     A close binary has a reflection effect that heats the substellar point
of each star.  Each star focuses the radiation from its companion on to itself.  
The radiative acceleration between the stars decreases the gravitational 
force pulling them together.  Both stars become more luminous and the envelope 
structure of each star responds to the additional energy with flows and mixing
and winds.  Tides become more complicated. 
\vfill
\eject
     Since gravitational force is so much simpler than electromagnetic force,
i.e. the cellular rules for gravity are simple compared to the cellular rules 
for charge, I see no need to postulate the existence of gravitational radiation. 

\vs
\vs
\centerline {Cosmological Redshift}
     A particle that travels any distance interacts with the background 
particles that it passes, gravitationally, electromagnetically,
and through every kind of interaction possible.  The passing particle, 
throughout its light cone, participates in the determination of the cellular 
motion of the background particles.  The background particles in the future 
light cone participate in the determination of the cellular 
motion of the passing particle.  Energy from the passing particle ``heats" 
the background particles.  A moving particle loses energy.  Photons and 
neutrinos are redshifted.  Cosmic rays lose energy. The redshift is determined 
by integration along the path.  The redshift is not rigorously a measure of 
distance or of time, but it is roughly proportional to distance and to time 
if the background density is statistically uniform in space and time.  The 
cosmological redshift is actually a ``cellular" redshift.  The universe is 
not expanding.  

     Photons follow indeterminate, irreproducible paths.  In special 
relativity photon paths are used to measure relative space and time.  But 
in physical special relativity the paths are ``fuzzy" so measurements of 
relative space and time are ``fuzzy" as well.  Special relativity is a 
mathematical approximation to the real physics.

\vfill
\eject
\centerline {\bf FUNDAMENTAL PARTICLES}

     Particles have complicated cellular automaton rules that involve many 
space cells and perhaps past and future (i.e. time derivatives).   Each type 
of particle has its own set of rules.  Charge, spin, charm, strangeness, etc.,
are cellular automaton rules.

     If massy particles are made from massless particles, then particle
physics must be modified.  Electrons and protons must be composed of massless
particles.  Quarks must be massless.  Since a neutron can be made from a proton
and an electron, it follows that an electron is composed of an anti-up-quark,
a down-quark, and an electron neutrino. 

     The fundamental particles are the 
six quarks and the three neutrinos and their antiparticles as shown in 
the table.  They all have spin 1/2 so are fermions.  All other particles are 
composite.  [The name of $\tau$ is regularized to tauon.]

\vs
\centerline {Fundamental Massless Particles}
{\settabs 20 \columns
\+charge && low energy    & &&&&& medium energy  &&&&&&& high energy   \cr
\+ +2/3 &&   u &up & &   &&&  c &charm & &&&&&    t &top \cr
\+ +1/3 &&  \=d &anti-down & &&&&    \=s &anti-strange &&&&&&    \=b &anti-bottom\cr
\+ \ \ \ \ \ 0 && $\bar{\nu}_{e}$ &electron& &&&& $\bar{\nu}_{\mu}$ &muon &&&&&& $\bar\nu_{\tau}$ &tauon\cr
\+&&&\ \ \ \ \ anti-neutrino &&&&&&\ \ \ anti-neutrino &&&&&&&\ \ anti-neutrino\cr
\+ \ \ \ \ \ 0 && $\nu_{e}$ &electron neutrino& &&&& $\nu_{\mu}$ &muon neutrino&&&&&& $\nu_{\tau}$ &tauon neutrino\cr
\+ \ -1/3 &&   d  &down & &&&& s &strange &&&&&& b &bottom\cr
\+ \ -2/3 &&   \=u &anti-up & &&&&    \=c &anti-charm &&&&&&        \=t &anti-top\cr
\+ \cr
}

\centerline {Simple Composite Particles}
{\settabs 10 \columns
\+  & particle    & &  & antiparticle  & & & energy(MeV)   \cr
\+  & p = uud  & & & \=p = \=u\=u\=d & & &  938.272 \cr
\+  & n = udd  & & & \=n = \=u\=d\=d & & &  939.566 \cr
\+  & $\pi^{+}$ = u\=d  & & & $\pi^{-}$ = \=ud & & & 139.568 \cr
\+  &  e$^{-}$ = \=ud$\nu_{e}$  & & &  e$^{+}$ = u\=d$\bar{\nu}_{e}$ & & & 0.511 \cr
\+  &  $\mu^{-}$ = \=ud$\nu_{\mu}$  & & &  $\mu^{+}$ = u\=d$\bar{\nu}_{\mu}$ & & & 105.658 \cr
\+  &  $\tau^{-}$ = \=ud$\nu_{\tau}$  & & &  $\tau^{+}$ = u\=d$\bar{\nu}_{\tau}$ & & & 1784.2 \cr
}
\vs
     All the fundamental particles are conserved.  All leptons contain a 
neutrino.  Pions are bosons because the quark spins cancel.  Fractional 
charges are probably one-dimensional and two-dimensional in some cellular 
automaton rules.  Free particles can be only three-dimensional so free quarks
cannot exist.  Here is a sampling of weak interactions with braces 
indicating virtual pairs:
\vs
n $\to$ n + $\{$e$^{+}$e$^{-}\}$ = udd + u\=d$\bar{\nu}_{e}$\=ud$\nu_{e}$ = 
uud + \=ud$\nu_{e}$ + $\bar{\nu}_{e}$ = p  + e$^{-}$ + $\bar{\nu}_{e}$

\vs
$\mu^{+} \to \mu^{+}$ + $\{$e$^{+}$e$^{-}\}$ = u\=d$\bar{\nu}_{\mu}$ +
u\=d$\bar{\nu}_{e}$\=ud$\nu_{e}$ = $\bar{\nu}_{\mu}$ + e$^{+}$ + $\nu_{e}$ 

\vs
$\mu^{-} \to \mu^{-}$ + $\{$e$^{+}$e$^{-}\}$ = \=ud$\nu_{\mu}$ +
u\=d$\bar{\nu}_{e}$\=ud$\nu_{e}$ = $\nu_{\mu}$ + e$^{-}$ + $\bar{\nu}_{e}$ 

\vs
$\pi^{+} \to \pi^{+}$ + $\{\mu^{+}\mu^{-}\}$ = u\=d +
u\=d$\bar{\nu}_{\mu}$\=ud$\nu_{\mu}$ =  $\mu^{+}$ + $\nu_{\mu}$ 

\vs
$\pi^{-} \to \pi^{-}$ + $\{\mu^{+}\mu^{-}\}$ = \=ud +
u\=d$\bar{\nu}_{\mu}$\=ud$\nu_{\mu}$ =  $\mu^{-}$ + $\bar{\nu}_{\mu}$ 

\vs
$\mu^{-}$ + p $\to$ \=ud$\nu_{\mu}$ + uud = $\nu_{\mu}$ + n

\vs
$\bar{\nu}_{e}$ + p $\to$ $\bar{\nu}_{e}$ + p + $\{$e$^{+}$e$^{-}\}$ = 
$\bar{\nu}_{e}$ + uud + u\=d$\bar{\nu}_{e}$\=ud$\nu_{e}$ = 
udd + u\=d$\bar{\nu}_{e}$ = n + e$^{+}$ 
 
\vs
p +p $\to$ p + p + $\{$e$^{+}$e$^{-}\}$ = uud + uud +  
u\=d$\bar{\nu}_{e}$\=ud$\nu_{e}$ 

\ \ \ \ \ \ \ \ \ \ \ \ \ \ \ \ \ \ \ \ \ \ \ \ \ \ \ \ \ \ \ \ \ \ \ = 
uud + udd + u\=d$\bar{\nu}_{e}$ + $\nu_{e}$ = pn + e$^{+}$ + $\nu_{e}$

\vs

If the lepton structure is equal to a pion plus a neutrino, what 
determines the mass?  Since mass is determined by internal motions
produced by cellular automaton rules, we can make an analogy of sorts 
(a gedanken astrophysics analogy (Kurucz 1992)) to the internal structure
of molecules expressed in partition functions.  A proton or neutron consists 
of three charged quarks that 
interact and execute internal motions.  Think of them as triatomic molecules 
that have huge partition functions because of all the modes of vibration.
Think of a pion as a diatomic molecule with a deep potential well with
many states but fewer states than the triatomic molecule so the mass of a 
pion is considerably less than that of a proton, $\pi$/p = 1/7.  Think of an 
electron as a diatomic molecule in which the electron neutrino somehow prevents 
the quarks from interacting more than one way.  There is only one level in a 
potential well so the electron mass is small, e/p = 1/1836 and e/$\pi$ = 1/280.  
Since the quark motion is localized in a small volume, an electron is 
a small coulomb target.  In the muon, the mu neutrino has some other rule that
does not strongly block the quark interaction.  The muon mass is reduced only 
slightly from that of a pion, $\mu/\pi$ = 3/4 and $\mu$/p = 1/9.  In the tauon, 
the tauon neutrino must enhance the quark interaction, perhaps by keeping the 
quarks from moving far apart.  It would act like a high barrier hump on the 
diatomic potential to produce more high-statistical-weight bound levels, so 
the tauon mass is much greater than the pion mass, $\tau/\pi$ = 13, and even 
larger than the proton mass, $\tau$/p = 2.
\vs
Since electrons are composite particles, they no longer
present singularity or normalization problems.  In low energy atomic physics
electrons and nucleons have impenetrable substructure.  

\vs
\vs
\centerline {\bf BLACK HOLES}
Black holes are easy to explain without singularities.  Adding mass to a
neutron star causes the neutrons to collapse into boson di-neutrons with 
spins anti-parallel,  n + n = udd + udd $\to$ uddudd = di-n.  The neutron 
star becomes a di-neutron star.  This can be a gradual transformation, not 
a catastrophe.  If matter is added slowly, the neutron star becomes an 
invisible black hole and continues to grow until the fermion nature of the 
quarks limits the compression.

Continuing to add mass to a di-neutron star causes the quarks within the 
di-neutrons to pair with spins opposed so they become di-quark bosons with 
0 spin, di-n = uddudd $\to$ uu-dd-dd = di-q-di-n.  The di-neutron star 
becomes a di-quark-di-neutron star which is a super-massive black hole. 

     A cellular automaton has to have a density limit to prevent overloading
the ``computation" in a small volume.  The simplest cutoff is to make 
gravity repulsive at high density to automatically blow apart dense
concentrations.  This could be built into the cellular rules for each particle.
\vs
\vs
\centerline {\bf ANTIMATTER}
There is a missing anti-matter problem if this universe began in a Big Bang
of radiation.  Starting with radiation implies that all primordial particles 
were made by pair production as the universe cooled.  If pair production
does not dominate, then matter and antimatter do not have to balance.

The idea of a di-quark-di-neutron black hole suggests that our universe 
``started" from a collection of ultra-massive black holes (UMBHs) 
statistically uniformly distributed throughout the cellular automaton.
For example, the initial state might be $10^{12}$ to $10^{13}$ 
$10^{13}$-to-$10^{14}$-solar-mass black holes with average separation 
less than a megaparsec.

At the first tick the density in the ultra-massive black holes exceded
the density cutoff so gravity was repulsive.  The ultra-massive black
holes expanded at sub-light speed.  The di-quark-di-neutrons expanded and 
became di-neutrons.  The di-neutrons
expanded and became neutrons.  The neutrons expanded and became protons and
electrons and anti-electron-neutrinos, and deuterons, and alpha particles, 
etc.  The initial number of neutrons was fixed.  The proton number, the
electron number, and the anti-electron-neutrino number are equal.  Subsequent
pair production does not affect the baryon (neutron+proton) total.  There is
no antimatter problem.  The universe is fundamentally biased toward matter.

The statistical equilibrium and the formation of nuclei led to different 
abundances, including heavier nuclei, and different properties than we are 
used to from a Big Bang prediction.

The expanding material from each ultra-massive black hole collided with the 
material from its neighbors and formed a pattern of density 
perturbations at galaxy scale, at globular-cluster scale, and at 
massive-Population-III-star scale.  The original locations of the ultra-massive 
black holes became regions of low density.  The perturbations evolved into 
the universe as we know it, including the background radiation.  I have 
described that evolution in my paper on radiatively-driven cosmology 
(Kurucz 2000).  I will produce an updated version.

\vfill
\eject

\centerline {\bf THE CELLULAR AUTOMATON UNIVERSE}

     The cellular automaton has a fixed number of cells, say a 
three-dimensional modular space cube with a large prime number K for the 
modulus.  Time has the same modulus. If a plausible minimum time in particle 
physics is $\sim$10$^{-28}$s,  the cell size would be $\sim$3$\times$10$^{-18}$ 
cm, about 10000 cells in a classical electron radius.   The cellular 
automaton is complete in itself.  If the modulus is, say, twice the age of 
the oldest stars, 30 billion years, K$\sim$10$^{46}$.  The number of 
cells, the largest physical number, is K$^{3}$ $\sim$10$^{138}$.
Each cell appears to be the center of the universe.

     Each type of particle has a set of rules that determine its motion.  
The occupation of each cell at each tick is determined by considering the 
occupation of that cell and all the cells neighboring in space and time.  
The definition of neighbor is a fundamental property of the cellular automaton.   
If the cellular grid is three dimensional, a cell can be defined to have 26 
neighbors, or 124 neighbors, or in other patterns that are considered in the 
cellular rules that determine particle motion. It can have neighbors in the 
past and neighbors in the future (as a predictor-corrector).  I suggest that
the cellular automaton has coordinates Nx, Ny, Nz, and Nt and that a particle 
can move in 26 ways, dNx and dNy and dNz = 0, $\pm$1, and dNt = +1.    
These coordinates are not the local x, y, z, and t coordinates of our 
experience in special relativity.

     The rule for momentum might be that a particle keeps track of and 
updates target coordinates Tx, Ty, Tz and tries to move to its target.  A 
particle would be a vector.  Interactions with other particles modify the 
choice of neighbor and the target coordinates.  

     A particle keeps track of its own energy with a counter Ne that has
the same range as the coordinates, K$\sim$10$^{46}$.  At each step it can
trade energy, increase or decrease or no change, with particles that it
can interact with.  In general a particle loses energy through interactions
when it travels over significant distances.

     The actual position and energy of a particle are unknowable.  They can
change $\sim$10$^{19}$ times per nanosecond.  There is no way to measure 
absolute position or time or energy, Nx, Ny, Nz, Nt, Ne.  Photons follow 
indeterminate, irreproducible paths.  In special relativity photon paths 
are used to measure relative space and time.  But in physical special 
relativity the paths are ``fuzzy" so measurements of relative space and 
time are ``fuzzy" as well.  Special relativity is a mathematical approximation 
to the real ``fuzzy" physics that makes the universe interesting.  

     Since integration over neighboring cells determines motion, and since 
the neighboring cells themselves also change at every tick, there is no 
time-reversal.  Physics cannot be run in reverse.  

     Particles and anti-particles have the same gravity rule: 
move in the direction that increases overlap with other particles.  
Above the density cutoff particles move in the direction that decreases 
their overlap.  

     Paradoxes do not fare well in a finite cellular automaton.  The 
solution to Olber's paradox is that the sky is dark because there is 
not enough light.

     Particles eventually circumnavigate the universe.  The cellular
automaton described here could run forever unless it learns to modify 
itself.

\vfill
\eject
     This work was supported, in part, by many NASA grants from 1974
through 2005.
\vs
\vs
\cl{References}
\ni Boulet, D.L. 1991. {\it Methods of Orbit Determination for the Microcomputer}

Richmond, Va.: Willmann-Bell.

\ni Kurucz, R.L. 1992. {Comments on Astrophysics}, 16, 1-16.

\ni Kurucz, R.L. 2000. {Radiatively-Driven Cosmology} arxiv:astro-ph/0003381.

\ni Lebretton, Y. and Dappen, W. 1988.  In {\it Seismology of the Sun and 
Sun-Like Stars}, 

ed. E.J. Rolfe, ESA SP-286, pp. 661-664.

\ni Misner, C.W., Thorne, K.S., and Wheeler, J.A. 1973.  {\it Gravitation}, New York:

Freeman.

\ni Resnick, R. 1968. {\it Introduction to Special Relativity}, New York: Wiley.

\ni Robertson, D.S., Carter, W.E., and Dillinger, W.H. 1991.  Nature 349, 768-770.

\ni Schubert, G., Ross, M.N., Stevenson, D.J., and Spohn, T. 1988.  In {\it Mercury}, ed. 

F. Vilas, C.R. Chapman, and M.S. Matthews, Tucson: University of Arizona 

Press, pp. 429-460.   

\ni Shapiro, I.I., Ash, M.E., Campbell, D.B. Dyce, R.B., Ingalls, R.P., Jurgens, R.F., 

and Pettengill, G.H. 1971.  Phys. Rev. Lett, 26, 1132-1135.

\ni Shapiro, I.I., Pettengill, G.H. Ash, M.E., Ingalls, R.P., Campbell, D.B., and  

Dyce, R.B. 1972.  Phys. Rev. Lett, 28, 1594-1597.

\ni Weisberg, J.M. and Taylor, J.H. 1984.  Phys. Rev. Lett. 52, 1348-1350.

\ni Wolfram, S. 1994. {\it Cellular Automata and Complexity}, Reading, Mass.: 

Addison-Wesley.

\vfill
\eject
\end